\documentstyle[10pt,epsf]{elsart}
\journal{Surface Science}
\begin{document}
\begin{frontmatter}
%..............................................................................

\title{Ab-initio study of SrTiO$_3$ surfaces}

\author{J.~Padilla and David Vanderbilt}

\address{Department of Physics and Astronomy, 
Rutgers University, 136 Frelinghuysen Road, Piscataway, NJ 08854-8019}

\date{February 18, 1998}
%..............................................................................

\begin{abstract}

We present first-principles total-energy calculations of (001)
surfaces of SrTiO$_3$.  Both SrO-terminated and TiO$_2$-terminated
surfaces are considered, and the results are compared with previous
calculations for BaTiO$_3$ surfaces.  The major differences are in
the details of the relaxed surface structures. Our calculations
argue against the existence of a large ferroelectric relaxation
in the surface layer, as had been previously proposed.  We do
find some indications of a weak surface ferroelectric instability,
but so weak as to be easily destroyed by thermal fluctuations
except perhaps at quite low temperatures.  We also compute
surface relaxation energies and surface electronic band structures,
obtaining results that are generally similar to those for
BaTiO$_3$.
\end{abstract}
%
% \begin{keyword}
% Density functional calculations; Surface relaxation and
% reconstruction; Surface energy; Dielectric phenomena
% \end{keyword}
\end{frontmatter}
%
%\pacs{77.80.-e, 68.35.Bs, 68.35.Md, 73.20.At, 68.35.-p}
%....................................................................

\section{Introduction}

The cubic perovskites are an important class of materials that are
of particular interest because of the variety of structural phase
transitions that they display \cite{line}.  The structural
instabilities may be of ferroelectric (FE) character, as for
BaTiO$_3$, or of antiferrodistortive character (involving
rotation of oxygen octahedra), as for SrTiO$_3$.  Specifically,
SrTiO$_3$ adopts a paraelectric simple cubic perovskite structure
above $T_c$=105K, but transforms to a tetragonal antiferrodistortive
structure
below $T_c$; and while it shows strong signs of FE fluctuations at
very low temperature, it evidently remains paraelectric down to
zero temperature.

The (001) and (111) surfaces of cubic perovskites have been most
investigated experimentally \cite{cox}.  For a II-IV
perovskite such as SrTiO$_3$, there are two possible nonpolar (001)
surface terminations: a SrO-terminated surface (type-I), and a
TiO$_2$-terminated surface (type-II).  On the other hand, the (111)
surfaces are polar, and therefore presumably much less stable.
Here we focus on the (001) surfaces of SrTiO$_3$.

Interest in the surface properties of SrTiO$_3$ arises because of
the catalytic properties of these surfaces \cite{tom}, and because
of the common use of SrTiO$_3$ as a substrate for epitaxial growth
of high-T$_c$ superconductors (such as YBa$_2$Cu$_3$O$_7$) \cite{chau}
and other oxides.

Theoretical studies of these surface have been
numerous.  Wolfram and coauthors \cite{wolf1}, using a linear
combination of atomic orbitals approach, predicted mid-gap surface states
for SrTiO$_3$, in disagreement with experimental investigations
\cite{powell,henrich}.  Tsukada {\it et al.} \cite{suka} employed the
DV X$\alpha$ cluster method to study SrTiO$_3$ surfaces, finding no
mid-gap surface states.  However cluster methods are not very
suitable for high-accuracy calculations of relaxations and
electronic states on infinite surfaces, underlining the need for
the application of more accurate, self-consistent techniques.
First-principles density-functional calculations have been very
successful in the study of bulk perovskites \cite{zhong1,wz2,cohen,singh},
and more recently there have been similar calculations for perovskite
surfaces \cite{padilla,cohen1,cohen2,kimura}.  In particular, Kimura 
{\it et al.} \cite{kimura} applied the same method as in this paper,
the plane-wave ultrasoft-pseudopotential method \cite{vand2}, to the
study of (001) surface of SrTiO$_3$, with and without
oxygen vacancies at the surface.  The main difference between that report
and the present work is that we fully relax the atomic coordinates
in the slab.  Also,
we analyze the possible existence of a FE surface layer for the 
SrO surface \cite{ravik}, as has been suggested previously.

On the experimental side, the study of these surfaces is
complicated by the presence of surface defects \cite{cord}, making
it difficult to verify the surface stoichiometry.  The results also
tend to depend upon the surface treatment \cite{hikita}.  Low-energy
electron diffraction (LEED) and reflection high-energy
electron diffraction (RHEED) studies of SrTiO$_3$ surfaces have
been reported \cite{hikita,bickel}.  These show no reconstruction of
the surface layer. In contrast, photoelectron spectroscopy
\cite{henrich1} and scanning tunneling microscopy \cite{matsu}
observations show different reconstructions for the reduced
surfaces.  The absence of in-gap surface states has also been
shown \cite{powell}.

\section{Theoretical approach}

As in previous work on BaTiO$_3$ surfaces \cite{padilla}, we employ
here a self-consistent pseudopotential technique in which the
valence electron wavefunctions are obtained by minimizing the
Kohn-Sham total-energy functional using a conjugate-gradient
technique \cite{king}.  The exchange-correlation potential is
treated within the local-density approximation (LDA) in the
Ceperley-Alder form \cite{ceper}. Vanderbilt ultrasoft
pseudopotentials \cite{vand2} are used to avoid norm-conservation
constraints, thus permitting the use of a small plane-wave cutoff
of 25 Ry, in spite of the fact of that we are dealing with
first-row and transition metal atoms. Such a cutoff has
previously been shown to be adequate in the bulk \cite{king}. The
forces on each ion were relaxed to less than $0.02$ eV/\AA\ using a
modified Broyden scheme \cite{vand1}.

Our calculations are carried out in a periodic slab geometry.  For
the type-I (SrO terminated) surface, the slab contains 17 atoms (4
SrO layers and 3 TiO$_2$ layers).  Similarly, the type-II (TiO$_2$
terminated) slab contains 18 atoms (4 TiO$_2$ layers and 3 SrO
layers).  In both cases, the slabs were thus three lattice
constants thick; the vacuum region was two lattice constants
thick.  The $z$-axis is taken as normal to the surface. The
calculations were done with a (6,6,2) Monkhorst-Pack
mesh \cite{mesh}.  This k-point set produced results of very
good accuracy. The structure was set up using our theoretical
lattice constant of 3.86 \AA, which is about 1\% smaller than the
experimental one; this underestimation is typical of LDA
calculations.  For further details concerning our method and its
accuracy, we direct the reader to our earlier paper \cite{padilla}.

Below 105K, SrTiO$_3$ transforms to a tetragonal antiferrodistortive
(AFD) state in which the oxygen octahedra rotate about a
$\langle$001$\rangle$ axis in opposite directions in alternate unit
cells \cite{vanderbilt}.  However, these rotations are typically
small; even at $T=0$ they amount to only $\sim 3^\circ$.
Since we are mainly interested in comparing with room-temperature
experiments, and since the static AFD distortions have already
disappeared at 105K (well below room temperature), we have not
included them in our calculations.  On the other hand, we do want to
consider the possible presence of a FE surface layer, as suggest in
Ref.\ \cite{ravik}.

We thus choose symmetries as follows.  Most of the calculations were
carried out using a``full set'' of symmetries, consisting of $M_x$,
$M_y$, and $M_z$ mirror symmetries (the latter being a reflection
through the center plane of the slab), as well as $1\times1$
translational symmetries parallel to the surface.  The full set of
symmetries prevents the occurrence of FE as well as AFD distortions.
Then, some further calculations were carried out using a ``reduced set''
of symmetries, identical to the full set except that $M_x$ symmetry
is allowed to be broken.  The reduced set of symmetries, while still
suppressing the AFD instability, allows the surface to develop a FE
distortion along $\hat x$ if it should turn out to be energetically
favorable.

\section{Surface relaxations}

\begin{table}
\caption{Equilibrium atomic displacements (relative to ideal positions)
for the SrO- and TiO$_2$-terminated surfaces, when no symmetry-breaking
distortions are allowed.  Units are at the theoretical 
lattice constant ($a=3.86$ \AA).}  
\vskip 0.3truecm
\begin{tabular}{llrlr}
\hline
\hline
Layer & SrO Surface &$\delta z$ & TiO$_2$ Surface &$\delta z$  \\
\hline
1 & Sr            &-0.057  & Ti           &-0.034 \\
  & O$_{\rm I}$   &0.001   & O$_{\rm II}$ &-0.016 \\
2 & Ti            &0.012   & Sr           &0.025 \\
  & O$_{\rm II}$  &0.   & O$_{\rm I}$  &-0.005 \\
3 & Sr            &-0.012  & Ti           &-0.007 \\
  & O$_{\rm I}$   &-0.001  & O$_{\rm II}$ &-0.005 \\
\hline
\hline
\end{tabular}
\label{tab:slabs}
\vskip 0.5truecm
\end{table}

We begin by presenting the relaxed structure for each of
the two surface terminations, obtained by starting from the
ideal structure and then relaxing the atomic positions while
preserving the full set of symmetries. The
relaxed geometries are summarized in Table \ref{tab:slabs}.
(Coordinates are only listed for atoms in the top half of the slab,
$z\ge0$; the others are determined by $M_z$ mirror symmetry).  By
symmetry, there are no forces along $\hat x$ or $\hat y$.  To set
notation, oxygen atoms O$_{\rm I}$ are the ones lying in SrO planes
of the slab, while O$_{\rm II}$ refers to the oxygen atoms lying in
TiO$_2$ planes of the slab.  The layer numbering in Table
\ref{tab:slabs} is from the outermost layer inwards.

From Table \ref{tab:slabs}, we can see that
the largest relaxations are for the metal atoms in the surface
layer, -5.7\% and -3.4\% for the Sr-terminated and Ti-terminated
surfaces respectively.  The outward relaxation of the second-layer
Sr atom on the Ti-terminated surface is also noteworthy.  The
surface-layer oxygen atoms show almost no displacement on the
Sr-terminated surface.

Previous surface structural refinements have been carried out by
Bickel {et al.} \cite{bickel} and by Hikita, Hanada and Kudo
\cite{hikita} using (LEED) and (RHEED),
respectively.  In Table \ref{tab:relax}, we compare our
structural parameters with the two experimental ones.  Both
experimental groups assumed that the oxygen and metal atoms
remain coplanar in the second and third layers in order to
simplify the refinement procedure; since the scattering strength
of O is much smaller than that of Sr and Ti \cite{bickel}, our
theoretical comparison is made to the position of the subsurface metal
layers.  Thus, we define $\Delta d_{12}$ as the change (relative
to bulk) of the first interlayer spacing, as measured from the
surface to the subsurface metal $z$ coordinate, and similarly for
$\Delta d_{23}$.  The quantity $s$ measures the outward
displacement of the surface-layer oxygens relative to the
surface-layer metal atoms.

\begin{table}
\caption{Comparison of theoretical and experimental structural
parameters for symmetry-preserving surface relaxations.
$\Delta d_{12}$ and
$\Delta d_{23}$ are respectively the changes in interlayer spacing
for the first and second pair of layers, while $s$ measures the
outward displacement of the surface oxygens relative to the
first-layer metal atoms.  All quantities are in \AA.}
\vskip 0.3truecm
\begin{tabular}{lccc}
\hline
\hline
  & $s$ & $\Delta d_{12}$ & $\Delta d_{23}$ \\
\hline
SrO-terminated & & & \\
\quad Theory, present                  &  0.22 & -0.26 &  0.10 \\
\quad Theory, Ref.\ \cite{mackt} &  0.14 & -0.47 &       \\
\quad Expt., Ref.\ \cite{bickel} &  0.16 & -0.19 &  0.08 \\
\quad Expt., Ref.\ \cite{hikita} &  0.16 &  0.10 &  0.05 \\
TiO$_2$-terminated & & & \\
\quad Theory, present                  &  0.07 & -0.27 &  0.12 \\
\quad Theory, Ref.\ \cite{mackt} &  0.04 & -0.39 &       \\
\quad Expt., Ref.\ \cite{bickel} &  0.08 &  0.04 & -0.04 \\
\quad Expt., Ref.\ \cite{hikita} &  0.10 &  0.07 &  0.05 \\
\hline
\hline
\end{tabular}
\label{tab:relax}
\vskip 0.5truecm
\end{table}

It is evident that the agreement between the theory and the experimental
refinements is not very good.  This is not surprising in view of the fact
that the experimental structures are in poor agreement with each other
\cite{explan-disagree}.
Our values for the rumpling $s$ of the surface layer are in rough
qualitative agreement with the experimental ones, though somewhat
larger in magnitude.  However, we predict a substantial contraction of
the interlayer spacing $d_{12}$, while in most cases the experimental
refinement indicates an expansion instead.  The one exception is for
the Sr-terminated surface, where our theory is in reasonable agreement
with Ref.\ \cite{bickel}, although not with
Ref.\ \cite{hikita}.  The agreement for the second interlayer
spacing can be seen to be mixed.

The disagreement between the two sets of experimental numbers
\cite{explan-disagree} suggests that the experimental refinements
should perhaps not be taken too seriously.  The expected quality
of a LEED or RHEED refinement is not well established for a complicated
metal oxide surface such as this one.  In the work of Bickel {\it et al.},
\cite{bickel} the authors were not able to determine independently
the proportions of the surface exhibiting the SrO and TiO$_2$
terminations; they assumed that the two appear in equal proportions.
Refining the structural parameters for both simultaneously, they
then obtained an $R$-factor of 0.53.  While this was argued to
be ``acceptable in view of the complexity of the structure,'' it
nevertheless seems uncomfortably large.  In the work of Hikita
{\it et al.,} \cite{hikita} the surface was prepared in different
conditions in order to obtain SrO and TiO$_2$ terminations separately;
for these, $R$-factors of 0.28 and 0.26 were obtained, respectively.
While this would thus appear to be the more reliable experiment,
unfortunately it is in no better agreement with the theory than the
refinement of Ref.\ \cite{bickel}.  One possible problem
with both experimental refinements could be the arbitrary assumption
that there is no buckling in the second metal-oxygen plane.
Our results indicate that there is a substantial
buckling in that layer, especially in the case of the subsurface
SrO layer on the TiO$_2$-terminated surface.

Also included in Table \ref{tab:relax} are theoretical estimates
of Mackrodt \cite{mackt} using an interatomic potential based on
Kim-Gordon pair potentials.  This theory gives results that are
qualitatively similar to ours, although the pair-potential model
appears to overestimate the size of the interlayer relaxations
and underestimate the degree of surface rumpling.  The pair-potential
model also predicts a significant rumpling in the subsurface SrO
layer of the TiO$_2$-terminated surface, similar to what we reported
above in Table \ref{tab:slabs}.

\section{Testing for a ferroelectric monolayer}

Bulk SrTiO$_3$ is an incipient ferroelectric: it nearly becomes
ferroelectric at very low temperature, and is apparently prevented
from doing so only by quantum zero-point fluctuations \cite{qsr}.
In our work on surfaces of FE BaTiO$_3$, we found some tendency
for an enhancement of the FE order in the first few surface layers,
especially for the case of the TiO$_2$ termination.
Thus, it is intriguing to speculate that if a similar tendency exists
in SrTiO$_3$, it might lead to the formation of a FE surface layer
on top of a paraelectric bulk material, at least at low temperatures.
In fact, the presence of just this kind of FE monolayer at the surface 
has been predicted by Ravikumar, Wolf and Dravid \cite{ravik} for
the SrO termination of SrTiO$_3$.  However, this prediction is based on
an empirical interatomic potential developed for ionic systems, and
it is not clear how far such a pair-potential approach can be trusted
for deciding such a delicate question as the appearance of surface
FE order.  Thus, we have undertaken to check whether such a FE surface
phase might occur in the context of our ab-initio calculations.

Note that while Bickel {\it et al.} have discussed the rumpling
of the surface layer in terms of a ``ferroelectric relaxation''
normal to the surface \cite{bickel}, such a rumpling does not
break any symmetry of the ideal surface, and is not qualitatively
different from the rumpling observed on other oxide surfaces.
Thus, when we speak of a FE surface layer, we shall restrict
ourselves here to symmetry-breaking distortions, i.e., frozen-in
displacements {\it parallel} to the surface.  Indeed, Ravikumar
{\it et al.} predict enormous displacements of this kind for the
SrO-terminated surface: Sr and O shift by 0.43\AA\ and -0.33\AA,
respectively, along a $\langle100\rangle$ direction, with a
resulting reduction in surface energy of 0.11 eV/cell.  When the
primitive $1\times1$ periodicity of the surface cell is allowed
to be broken, they find an even lower-energy c2$\times$2 structure.

With this motivation, we have investigated carefully the stability
of our $1\times1$ surface with respect to displacements of the
surface atoms along the $x$ axis, using the reduced set of
symmetries as discussed in Section 2.  Starting from the structure
of Table \ref{tab:slabs}, an additional FE distortion along $\hat x$
is imposed, the value of this distortion being taken the same
as for bulk BaTiO$_3$ as computed in Ref.\ \cite{king}.  Then we
allow this structure to relax fully.  We did these simulations
for both kinds of termination, and for both theoretical and
experimental lattice constants.

For the case of experimental lattice constant, the resulting 
relaxed displacements along the $\hat x$ direction,
in units of lattice constant, are as follows.
For the SrO-terminated surface, the first- and third-layer Sr
displacements are 0.005 and 0.006; the second- and fourth-layer
Ti displacements are 0.005 and 0.008; and the oxygen displacements
are -0.007, -0.002, -0.006, and -0.004 in layers 1 through 4
respectively.
For the TiO$_2$-terminated surface, the first- and third-layer Ti
displacements are 0.017 and 0.016; the second- and fourth-layer
Sr displacements are 0.010 and 0.011; and the oxygen displacements
are -0.011, -0.008, -0.007, and -0.009 in layers 1 through 4
respectively.  The total energies lie below those 
of the high-symmetry structure
(i.e., no $\hat x$ distortion) by 0.001 and 0.026 eV for the SrO- and
TiO$_2$-terminated surfaces, respectively.  We see that
the final FE distortions left after the relaxations are indeed very
small, and are not much greater in the surface layer than in the
deeper layers.  When using the theoretical lattice constant instead
of the experimental one, we found only a very weak distortion of
energy 0.001 eV for the TiO$_2$-terminated case, and no observable
distortion for the SrO-terminated case.

These results indicate that there does exist the possibility of
a small FE surface distortion for both surface terminations of
SrTiO$_3$.  However, the double-well depth that we found for
the SrO surface is very much smaller than the one predicted by
Ravikumar {\it et al.}\cite{ravik}.  We actually find a somewhat
larger tendency for FE distortion for the TiO$_2$ surface,
but still the distortion amplitudes (maximum 0.07\AA)
and energy (0.013 eV/surface) are nearly an order of magnitude
smaller than the SrO-surface prediction of Ref.~\cite{ravik}.
Our predicted FE distortion is small enough that it may easily
be destroyed by thermal (or even quantum) fluctuations, and
it seems unlikely that it would appear at room temperature.
However, it seems possible that the FE distortion might survive
and be observable at very low temperatures.

\section{Surface energies}

We discuss next the surface energetics.  To compare energies of
surfaces having different stoichiometry, we follow the approach 
described in Ref.\ \cite{padilla}.  That is, we consider each
surface slab to be built from TiO$_2$ and $SrO$ units, and
compute the grand thermodynamic potential $F$ for each type of surface
as a function of the TiO$_2$ chemical potential $\mu_{\rm TiO_2}$.
The energies of the bulk crystals of SrO and TiO$_2$ were calculated
with the LDA using the same pseudopotentials, and the same pane-wave cut-off.
The final results are shown in Fig.~1.  It can be seen that both
surfaces have a comparable range of
thermodynamic stability, indicating that either termination
could be formed depending on whether growth occurs
in Sr-rich or Ti-rich conditions, with a very small preference for the
TiO$_2$ termination.

\begin{figure}
\epsfxsize=3.1 truein
\centerline{\epsfbox{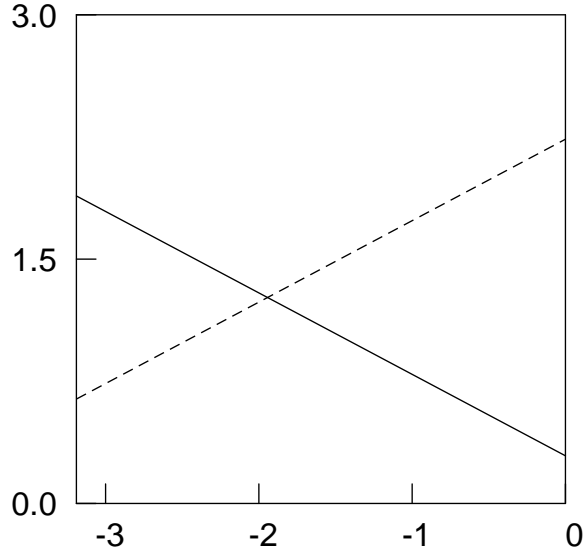}}
\caption{
Grand thermodynamic potential $F$ as a function of the chemical
potential $\mu_{\rm TiO_{\rm 2}}$,
for the two types of surfaces.
Dashed and solid lines correspond to type-I (SrO-terminated)
and type-II (TiO$_2$-terminated) surfaces respectively.
\label{fig1}}
\vskip 0.5truecm
\end{figure}

The average of $F$ for the two types of surface, which we shall
denote as $E_{\rm surf}$, is independent of $\mu_{\rm TiO_2}$.
Thus, this quantity is convenient for comparisons.  The value we
found for $E_{\rm surf}$ for the SrTiO$_3$ (001) surfaces is
1.26 eV per surface unit cell ($1358$ erg/cm$^2$).  This is very
similar to the corresponding value 1.24 eV that was obtained for
the BaTiO$_3$ (001) surfaces (in the cubic phase).

To compute the surface relaxation energy $E_{\rm relax}$,
we computed the average surface energy $E_{\rm unrel}$ for the
{\it unrelaxed} slabs (i.e., atoms in the ideal cubic
configuration), using the same k-point sampling as for
the relaxed systems.  We obtained $E_{\rm unrel}=1.44$ eV.
Thus, the relaxations account for $0.18$ eV of the surface energy
per surface unit cell, accounting for around 15\% of the
total surface energy.

\section{Surface electronic structure}

\begin{figure}
\epsfxsize=3.1 truein
\centerline{\epsfbox{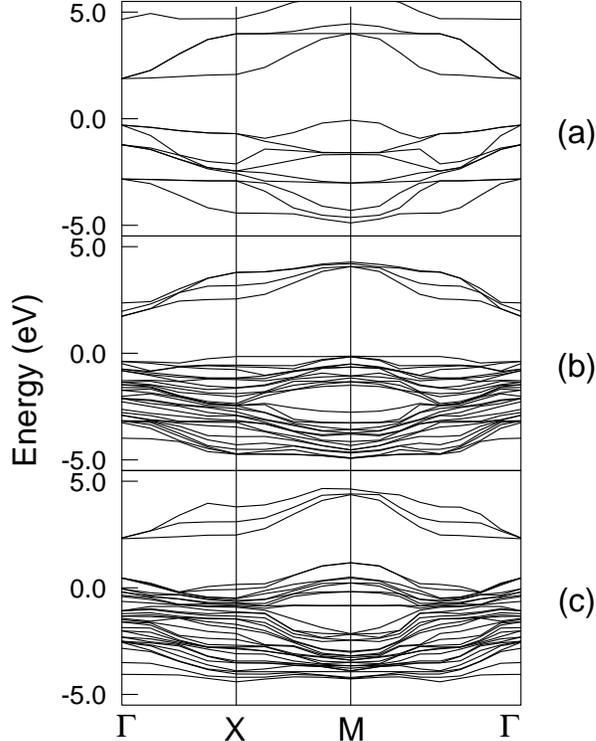}}
\caption{
Calculated electronic band structures for SrTiO$_3$.
(a) Surface-projected bulk band structure.
(b) SrO-terminated surface (slab I).
(c) TiO$_2$-terminated surface (slab II).
The zero of energy corresponds to the bulk valence band maximum.
Only the lowest few conduction bands are shown.
\label{fig2}}
\vskip 0.5truecm
\end{figure}

We focus now on the LDA calculated electronic structure for the
surfaces slabs. Although the LDA is well known to underestimate the
band gaps, we can be assured that the results given here are at least
qualitatively correct.
The band gap we obtained for bulk is SrTiO$_3$ 1.85 eV, to be compared with
the experimental value of 3.30 eV; this level of disagreement is
typical for the LDA. 

Not surprisingly, our results for the band structures of the surfaces
of SrTiO$_3$
are very similar to the case of BaTiO$_3$ surfaces \cite{padilla}. 
The computed LDA electronic energy band structure is given in 
Fig.~2 and the direct
energy gaps are given in Table \ref{tab:gaps}.  We see that the
band gap for the SrO surface almost does not change with
respect to the bulk value, and no in-gap state occurs. For the
TiO$_2$ surface, there is a substantial reduction of the band gap.
However, there are no deep-gap surface states,
in accord with experimental reports \cite{powell}.

From Fig.~2(c), we see (as is also true for BaTiO$_3$) \cite{padilla} 
that there is a tendency for valence-band states to intrude upwards
into the lower part of the band gap for this surface, especially
near the $M$ point of the surface Brillouin Zone, that is the top of
the valence band. In order to analyze the character of the
valence-band state at the $M$ point of the TiO$_2$ surface, the charge
density at this state was obtained. The results are very similar to
those shown in Fig.~4 of Ref.~\cite{padilla} for the case of
BaTiO$_3$. These states correspond to O $2p$ orbitals lying in the
surface plane and having very little hybridization with Ti $3d$ levels
because of the existence of four nodal planes [(100), (110), (010), and
($\bar 1$10)] intersecting at the Ti sites.

\begin{table}
\caption{Calculated electronic energy band gaps for the relaxed 
surface slabs (eV).}
\vskip 0.3truecm
\begin{tabular}{lc}
\hline
\hline
Slab             & Band gap\\
\hline  
Slab I           & 1.86  \\
Slab II          & 1.13  \\
Bulk             & 1.85  \\
\hline
\hline
\end{tabular}
\label{tab:gaps}
\vskip 0.5truecm
\end{table}

\section{Summary}

In summary, we have carried out LDA density-functional calculations of
SrO- and TiO$_2$-terminated (001) surfaces of SrTiO$_3$.  
By minimizing the Hellmann-Feynman forces on the atoms,
we obtained the optimal ionic positions.
Previous experimental LEED \cite{bickel} and RHEED \cite{hikita}
surface structure determinations are found to be significantly at
variance with our predictions, as well as with each other.

Our calculations do not support the existence of a large FE distortion
in the surface layer of the Sr-terminated surface, as had been previously
proposed theoretically \cite{ravik}.  While we cannot rule out
a weak FE surface instability, especially for the Ti-terminated
surface, our results indicate that it would be sufficiently weak so as
to be easily destroyed by thermal fluctuations except at quite low
temperatures.
The equilibrium state of the surface was obtained as a function
of the relative Sr and Ti stoichiometry.
The average energy of the Sr- and Ti-terminated surfaces was
found to be $\sim1360$ erg/cm$^2$, and the surface relaxation
energy was calculated to be $\sim190$ erg/cm$^2$.
In agreement with experiments, no deep-gap
surface level is found. However, as found for BaTiO$_3$, there is
a significant reduction of the electronic band gap of the
TiO$_2$ surface. This reduction arises from surface states in the
lower part of the bulk band gap which can be regarded
as resulting from the intrusion of valence-band states of O
$2p$ character, particularly in the vicinity of the M point
of the surface Brillouin zone.

\begin{ack}
The present work was supported by ONR grant N00014-97-1-0048
and NSF grant DMR-96-13648.

\end{ack}

%-----------------------------------------------------------------------

\begin{thebibliography}{99}

\bibitem{line} M. E. Lines and A.M. Glass, {\it Principles and
Applications of Ferroelectrics and Related Materials,} (Clarendon
Press, Oxford, 1977).

\bibitem{cox} V.E. Henrich and P.A. Cox, {\it The Surface Science
of Metal Oxides,} (Cambridge University Press, New York, 1994).

\bibitem{tom} M. Tomkiewicz and H. Fay, Appl. Phys. {\bf 18} (1979) 1.

\bibitem{chau} P. Chaudhari,  R. H. Koch, R. B. Laibowitz,
T. R. McGuire, and R. J. Gambino, Phys. Rev. Lett. {\bf 58}
(1987) 2684.

\bibitem{wolf1} T. Wolfram, E.A. Kraut, and F.J. Morin, Phys. Rev. B
{\bf 7} (1973) 1677.

\bibitem{powell} R.A. Powell and W.F. Spicer, Phys. Rev. B
{\bf 13} (1976) 2601.

\bibitem{henrich} V.E. Henrich, G. Dresselhaus, and H.J. Zeiger,
Bull. Am. Phys. Soc. Ser. II {\bf22} (1977) 364.

\bibitem{suka} M. Tsukada, C. Satoko, and H. Adachi, J. Phys. Soc. 
Jpn. {\bf 48} (1980) 200.

\bibitem{zhong1} W.~Zhong, R.~D.~King-Smith and D.~Vanderbilt, Phys.  
Rev. Lett.  {\bf 72} (1994) 3618.

\bibitem{wz2} W.~Zhong, D.~Vanderbilt, and K.M. Rabe, Phys. Rev.
Lett. {\bf73} (1994) 1861; Phys. Rev. B. {\bf 52} (1995).

\bibitem{cohen} R.E. Cohen and H. Krakauer, Phys. Rev. B
{\bf 42} (1990) 6416; Ferroelectrics {\bf 136} (1992) 65; 
R.E. Cohen, Nature {\bf 358} (1992) 136.

\bibitem{singh} D.J. Singh, Ferroelectrics {\bf 164} (1995) 143.

\bibitem{padilla} J. Padilla and D. Vanderbilt, Phys. Rev. B
{\bf 56} (1997) 1625.

\bibitem{cohen1} R.E. Cohen, J. Phys. Chem. Solids {\bf 57} (1996) 1393. 

\bibitem{cohen2} R.E. Cohen, Ferroelectrics {\bf 194} (1997) 323.

\bibitem{kimura} S. Kimura, J. Yamauchi, M. Tsukada, S. Watanabe, 
Phys. Rev. B {\bf 51} (1995) 11049.

\bibitem{vand2} D. Vanderbilt, Phys. Rev. B {\bf 41} (1990) 7892.

\bibitem{ravik} V. Ravikumar, D. Wolf, V. P. Dravid, Phys.  
Rev. Lett. {\bf 74} (1995) 960.

\bibitem{cord} B. Cord and R. Courths, Surf. Sci. {\bf 152/153} (1985) 1141.

\bibitem{hikita} T. Hikita, T. Hanada and M. Kudo, Surf. Sci. 
{\bf 287/288} (1993) 377.

\bibitem{bickel} N. Bickel, G. Schmidt, K. Heinz, K. Muller,
Phys. Rev. Lett. {\bf 62} (1989) 2009; Vacuum {\bf 41} (1990) 46.

\bibitem{henrich1} V.E. Henrich, G. Dresselhaus, and H.J. Zeiger,
Phys. Rev. B {\bf 17} (1978) 4908.

\bibitem{matsu} T. Matsumoto, H. Tanaka, K. Kogouchi, T. Kawai,
S. Kawai, Surf. Sci. {\bf 312} (1993) 21.

\bibitem{king} R.D. King-Smith and D. Vanderbilt, Phys. Rev. B {\bf
49} (1994) 5828.

\bibitem{ceper} D.M. Ceperley and B.J. Alder,
Phys. Rev. Lett. {\bf 45} (1980) 566.

\bibitem{vand1} D. Vanderbilt and S.G. Louie,
Phys. Rev. B {\bf 30} (1984) 6118.

\bibitem{mesh} H.J. Monkhorst and J.D. Pack, Phys. Rev. B {\bf 13}
(1976) 5188.

\bibitem{vanderbilt} D. Vanderbilt and W. Zhong, Phys.
Rev. Lett. {\bf 74} (1995) 2587.

\bibitem{mackt} W.C. Mackrodt, Phys. Chem. Min. {\bf 15} (1988) 228.

\bibitem{explan-disagree}
While a statement appears in Ref.\ \cite{hikita} to the effect
that their results are ``basically consistent'' with those of
Ref.\ \cite{bickel}, this statement is not supported by a
direct comparison of the two sets of results, as in Table \ref{tab:relax}.

\bibitem{qsr} W. Zhong and D. Vanderbilt,
Phys. Rev. B {\bf 53} (1996) 5047.

\end{thebibliography}
\end{document}